# 1000-Channel Integrated Optical Phased Array with 180° Field of View, High Resolution and High Scalability

YONG LIU, XIANSONG MENG, AND HAO HU*

*Department of Electrical and Photonics Engineering, Technical University of Denmark, 2800 Kgs. Lyngby, Denmark*
*huhao@dtu.dk ORCID: 0000-0002-8859-0986

**Abstract:** Optical phased array (OPA) is a promising technology for compact, solid-state beam steering, with applications ranging from free-space optical communication to LiDAR. However, simultaneously achieving a large field of view (FOV), high resolution, and low side-lobe level (SLL) remains a major challenge. Traditional OPAs face inherent limitations: they exhibit grating lobes when emitter spacing exceeds half the operating wavelength, while at half-wavelength spacing, significant crosstalk issues persist. Previously, we demonstrated a small-scale OPA that harnesses near-field interference and beamforming via a trapezoidal slab grating and a half-wavelength-pitch waveguide array to achieve a 180° FOV. However, its resolution was limited by the small channel count. In this work, we present a 1000-channel OPA that scales this architecture while addressing key challenges in waveguide crosstalk and control complexity. By optimizing waveguide routing, we minimize inter-channel coupling in the dense waveguide array. Additionally, we propose and demonstrate a passive matrix control scheme using 20 row and 50 column pulse-width modulation (PWM) signals to arbitrarily control 1000 thermo-optic phase shifters, significantly simplifying the electronic control and packaging. Our OPA achieves grating-lobe-free beam steering across a full 180° FOV, with a high resolution of $0.07° \times 0.17°$ and a minimum SLL of -18.7 dB at 0°. This large-scale, cost-effective chip-based OPA paves the way for next-generation high-resolution, wide-angle beam steering systems.

## 1. Introduction

Optical phased array (OPA) steers optical beams by adjusting the phase of an array of emitters, with a broad range of applications spanning free space optical communication [1], holographic imaging [2–4], bio-imaging [5–7], and light detection and ranging (LiDAR) [8–12]. As OPAs require no mechanical moving parts, they are considered as one of the most promising technologies for fast and reliable solid-state beam steering. These photonic integrated circuits (PICs), which integrate components such as beam splitters, phase shifters and grating antennas, can be fabricated on various integrated photonic platforms including silicon, silicon nitride, and lithium niobate [13–16]. The compactness and CMOS-compatible manufacturing of integrated OPAs could significantly reduce the cost of beam-steering systems while maintaining high performance.

Traditionally, OPA based beam steering relies on far-field interference among grating emitters. However, grating lobes emerge when the emitter pitch exceeds half the wavelength [17], while sub-wavelength spacing leads to severe crosstalk, significantly degrading beam quality. Non-uniform spacing between emitters can reduce the grating lobes by redistributing the power to side lobes [18,19]. However, these approaches compromise the emission area's filling factor, reducing the main lobe efficiency. To address these challenges, we previously demonstrated an integrated OPA combining a slab grating with a half-wavelength-pitch waveguide array, effectively eliminating grating lobes and minimizing crosstalk [20]. In this configuration, the width and routing geometry of the 64-channel waveguide array were optimized to minimize the coupling between adjacent waveguides. However, a key remaining question is whether this approach remains effective when scaling the channel count by an order of magnitude. In particular, the transition region between the phase shifter array and the half-wavelength-spaced waveguide array necessitates longer waveguide segments, which may increase coupling and pose new design challenges.

In addition, increasing the number of channels in OPAs also demands a proportional increase in electronics control. Typically, each phase shifter requires an independent control signal, often provided by a digital-to-analog converter (DAC) [1]. For OPAs with a small number of channels, these DACs can be connected to the chip via wire bonding. However, as the channel count grows, this approach necessitates more complex control circuitry and intricate packaging [21], significantly increasing the cost. Reducing the number of electrical components can simplify the design and lower overall costs. Several strategies have been proposed to share controlling electronics among multiple channels [22,23], however these methods are highly susceptible to fabrication imperfections, severely limiting their scalability. An alternative solution lies in row-column addressing using pulse width modulation (PWM), a technique widely adopted in large-scale light-emitting diode (LED) dimming [24], which enables control of millions of pixels with relatively few signal lines. Similar PWM controlling method has adopted in a diode based thermal optical (TO) phase shifter matrix and controlled the $N \times M$ phase shifters with only $N+M$ signals [25]. A 128-channel OPA has also been demonstrated by PWM control [26] with doped TO phase shifters. We will bring this method to a larger scale of 1000 channels with passive TO phase shifters only.

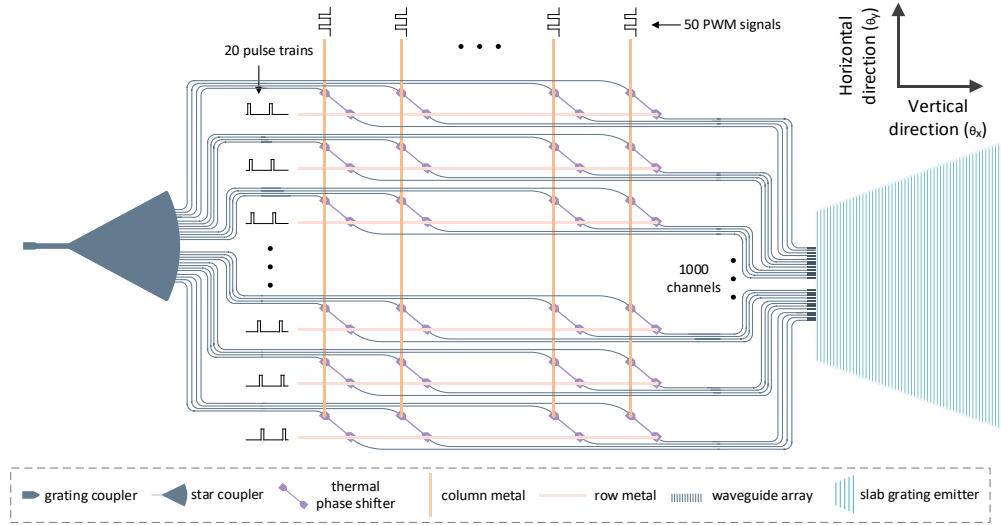

Fig. 1. Schematic of the Chip-scale optical phased array.

In this work, we present and experimentally demonstrate a 1000-channel OPA comprising an input grating coupler, a 1-to-1000 star coupler, 1000 TO phase shifters, a half-wavelength-pitch waveguide array, and a slab grating, as shown in Fig. 1. The TO phase shifters can be arbitrarily controlled using only 20 row and 50 column PWM signals by calculating effective power of each phase shifter and analytically solving the corresponding equations. This large-scale integrated OPA achieves a 180° field of view (FOV) while maintaining high beam quality, with a side lobe level (SLL) as low as -18.7 dB at 0°. The width and routing geometry of the 1000-channel waveguide array are carefully optimized to minimize inter-channel coupling within the densely packed half-wavelength-pitch array. The star coupler distributes the input light into all channels, generating a Gaussian amplitude distribution to suppress side lobes in the far field. Furthermore, the combination of the half-wavelength-pitch waveguide array and the trapezoidal slab grating enables near-field interference and beamforming within the slab region, followed by far-field emission, resulting in grating-lobe-free beam steering across the entire 180° FOV.

## 2. Waveguide array and slab grating design for OPA at large scale

Due to the physical dimensions of the TO phase shifters, the 1000 waveguide channels initially fan out to a large spacing (~7.5um) to connect with the phase shifters before converging to the half-wavelength pitch (0.775 µm) at the slab grating interface, as shown in Fig. 1. The routing of the waveguide array is optimized to minimize crosstalk. If the spacing between adjacent waveguides exceeds 2 µm, crosstalk is considered as negligible. However, where the spacing falls below 2 µm, this section of the waveguide array should be carefully designed to minimize crosstalk. Fig. 2(A) shows a close-up view of the waveguide array, where the spacing between adjacent waveguides gradually decreases from 2 µm to the half-wavelength pitch. The bending radii of the waveguides are optimized to ensure that the waveguide length of each channel in this section is as equal and as short as possible, therefore minimizing and uniformly distributing crosstalk across the array. In the outer sections of the array (dark-colored region), the waveguides follow a series of quarter circles with linearly increasing radii. In the central section (light-colored region), the waveguides bend with a calculated, increasing radius to maintain equal lengths and designated spacing throughout the structure. The calculated waveguide length of the array increases nearly linearly with the channel count, reaching 632 µm for our 1000-channel OPA. Additionally, the waveguide array is segmented into superlattice cells, each consisting of waveguides with widths of 560 nm, 380 nm, 580 nm and 400 nm to further reduce crosstalk [27]. It's worth noting that the superlattice only start after the waveguides converging to the half-wavelength pitch. For a waveguide array with a length of 626 µm and a four superlattice cells configuration, finite-difference time-domain (FDTD) simulations indicate that the crosstalk between adjacent waveguides can be as low as -35 dB. Notably, the crosstalk remains at -35 dB even when the waveguide length in this section increases to 2000 µm, which is the maximal length we can simulate in FDTD (Fig. 2(B)). This corresponds to an approximately 3200-channel OPA with a half-wavelength pitch. On the other hand, variations in waveguide width complicate compensation of the path length difference between channels, which is necessary for phase calibration. In this device, path length compensation is not applied, but it could be achieved by adding equivalent length of waveguides with varying widths to all channels, although at the cost of increased circuit size.

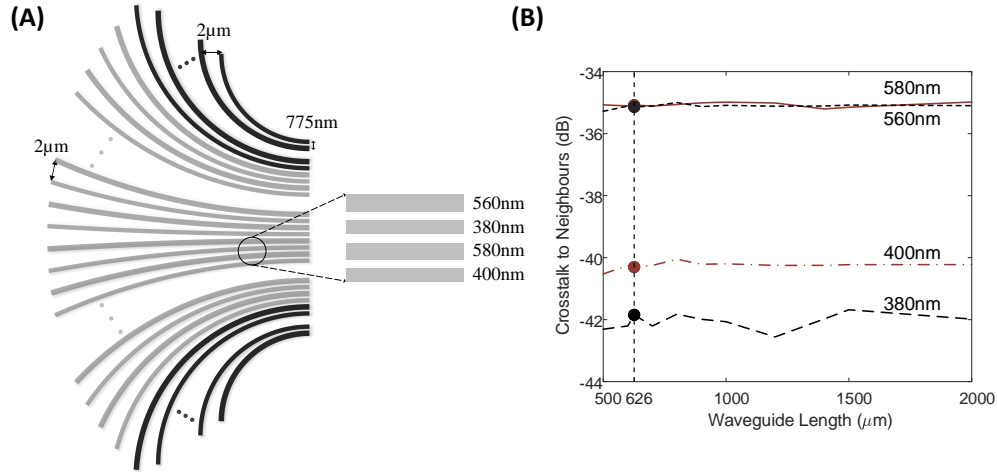

Fig. 2. Waveguide array. (A) The zoom-in structure of waveguide array. (B) Simulated crosstalk of the waveguide array with four waveguide widths versus the waveguide length.

The slab grating, positioned 200 µm after the waveguide array, directs the in-plane beam out of the chip. Unlike waveguide gratings, the slab grating has no confinement in lateral direction, therefore the near-field beams emitted from the end of waveguide array interfere and form a steerable beam within the chip plane before being redirected. The slab grating is designed with a trapezoidal shape to effectively contain the near-field beam. The grating pitch and duty cycle were optimized through simulations to be 560 nm and 50%, respectively, enabling near-vertical emission at an input wavelength of 1550 nm. Fabrication of the slab grating involves a shallow etch of 10 nm into the silicon layer, with a total grating length of 4 mm. The extended grating length ensures near-complete optical power emission and enhances the fabrication tolerance for etch-depth variations. The long, weak grating facilitates a large emission aperture in vertical direction marked in Fig. 1.

The combination of the half-wavelength pitch waveguide array and the slab grating enables grating-lobe-free emission in the far field and achieves a 180° FOV [20]. This design offers significant advantages over conventional waveguide array grating based OPAs. First, the length of the waveguide array with sub-wavelength spacing is substantially shorter than the waveguide array grating length required in conventional OPAs, minimizing the region vulnerable to inter-channel crosstalk. Second, the waveguide width in our OPA can be freely optimized to suppress crosstalk through induced phase mismatching. However, this method is not applicable to waveguide array gratings, as width variations would result in different emission angles along the grating.

## 3. Phase tuning by PWM signals

The TO phase shifters employ a triple-folded parallel waveguides to maximize heat utilization from the heater. To further reduce power consumption, the waveguides are suspended from the silicon substrate. Fig. 4(F) shows the top-view image of the heaters, where two trenches parallel to the waveguides are etched through the bottom SiO2 layer down to the silicon substrate using isotropic etching. The silicon substrate is undercut to a depth of 30 μm to achieve sufficient thermal isolation. The power required for a π phase shift, measured separately using another sample, is 0.14 mW/π, representing a 37-fold improvement in efficiency compared to phase shifters without undercut. The undercut, on the other hand, has reduced the working speed of the phase shifters, and we expect the response time is on the order of 200-300 μs [28]. The heaters are arranged in a 20 × 50 matrix and addressed through metal lines organized in rows and columns. PWM signals are connected to the 1000 heaters via 70 pads, significantly reducing the packaging footprint and complexity of wire bonding.

Figure 3(A) shows the PWM control architecture. The array is driven by $n$ sequential row pulse trains and $m$ column PWM channels. When addressing heaters, individual row and column lines are either driven high or grounded. Resistors that share the same potential at both terminals are treated as being in parallel, which permits the network to be reduced to the equivalent circuit shown in Fig. 3(B), with $m_0$ grounded inputs and $m_1$ high inputs in rows, $n_0$ grounded inputs and $n_1$ high inputs in columns.

For clarity we assume each FPGA I/O port has an output resistance $R_0$ and every heater channel has the same resistance R (measured to be ~500 Ω for the OPA device). The four possible terminal-voltage states of any equivalent resistor are denoted $U_k$ ($k = 1,2,3,4$), with potential at two ends of 00, 0V, V0, VV, respectively. (0 denotes grounded, V denotes high) The voltage across a resistor, and consequently its instantaneous power, is governed by the number of inputs ($m_0, m_1, n_0, n_1$) that place it in each possible state as the row and column signals are modulated these counts change in time and each resistor therefore cycles through the four voltage states with different durations.

If the PWM frequency is substantially higher than the thermal–optical (TO) response frequency of the heaters, the heater responds to the time-averaged electrical power. Under this quasi-steady assumption, the effective electrical power delivered to the heater at position ($i$th row, $j$th column), is

$$P_{ij} = \frac{1}{mR} \sum_{k=1}^{4} \sum \overrightarrow{\Delta d_k} \cdot \overrightarrow{U_k^2} \tag{1}$$

where $\overrightarrow{U_k^2}$ is the time-average of the squared voltage in state $k$, and $\overrightarrow{\Delta d_k}$ is the (time-averaged) weighting factor for state $k$ that depends on the duty-cycle pattern $d_{ij}$ applied on the column lines. A full derivation and the explicit form of $\overrightarrow{\Delta d_k}$ are provided in the supplementary information.

In the implemented scheme, the heater rows are activated one at a time in sequence, using pulse trains of fixed period. During the active period of each row, all the column drivers apply one cycle of pulse-width-modulated (PWM) signal. Each PWM signal has a duty cycle $d_{ij}$, where the index $i$ refers to the row being addressed and $j$ to the column — see Fig. 3(A). After all $m$ rows have been addressed once, the sequence repeats. This full cycle is defined as the control period. Equation (1) shows that the effective heating power of each element depends on the $m$ duty cycles applied across the $n$ columns, giving $m \times n$ independent duty-cycle parameters in total.

Due to the row–column multiplexing, however, the array requires only $m + n$ physical drive signals (instead of one per heater). Despite this signal reduction, the $m \times n$ governing equations can, in principle, be solved analytically, enabling arbitrary power distributions across the entire array. The control rate is set by the length of the row pulse trains (i.e., the control period) and therefore scales with the row number $m$. To balance the requirement that the control be faster than the heaters' TO response while keeping the number of I/O ports practical (limited here to ≤100), we selected $m$ to be 20.

Fig. 3(D) presents oscilloscope measurements of the FPGA-generated signals. In each IO of FPGA, the high level voltage is ~2.5V, and maximal current supported is 24mA. The unloaded FPGA outputs are shown in orange and thick traces. The row outputs form sequential pulse trains in which only one row is driven high at a time; the column outputs are PWM waveforms with varying duty cycles. To reduce switching activity, alternate PWM column signals are polarity inverted, which halves the total number of rising and falling edges compared to a non-inverted scheme (Fig. 3(C)), improving signal integrity.

The blue, thin traces in Fig. 3(D) show the same FPGA outputs under load (connected to the heater array). The transitions remain synchronized with the unloaded signals, but the previously flat high and low levels develop finite slopes. This is attributable to the distributed loading: each input port contributes to the voltage across multiple resistors (as reflected in Equation (1)), which effectively raises the low-level baseline and lowers the high-level amplitude at the outputs.

Finally, column duty cycles are tuned to achieve the desired phase shifts using a univariate search algorithm [29], which is the same optimization routine we applied previously to the 64-channel OPA system [20].

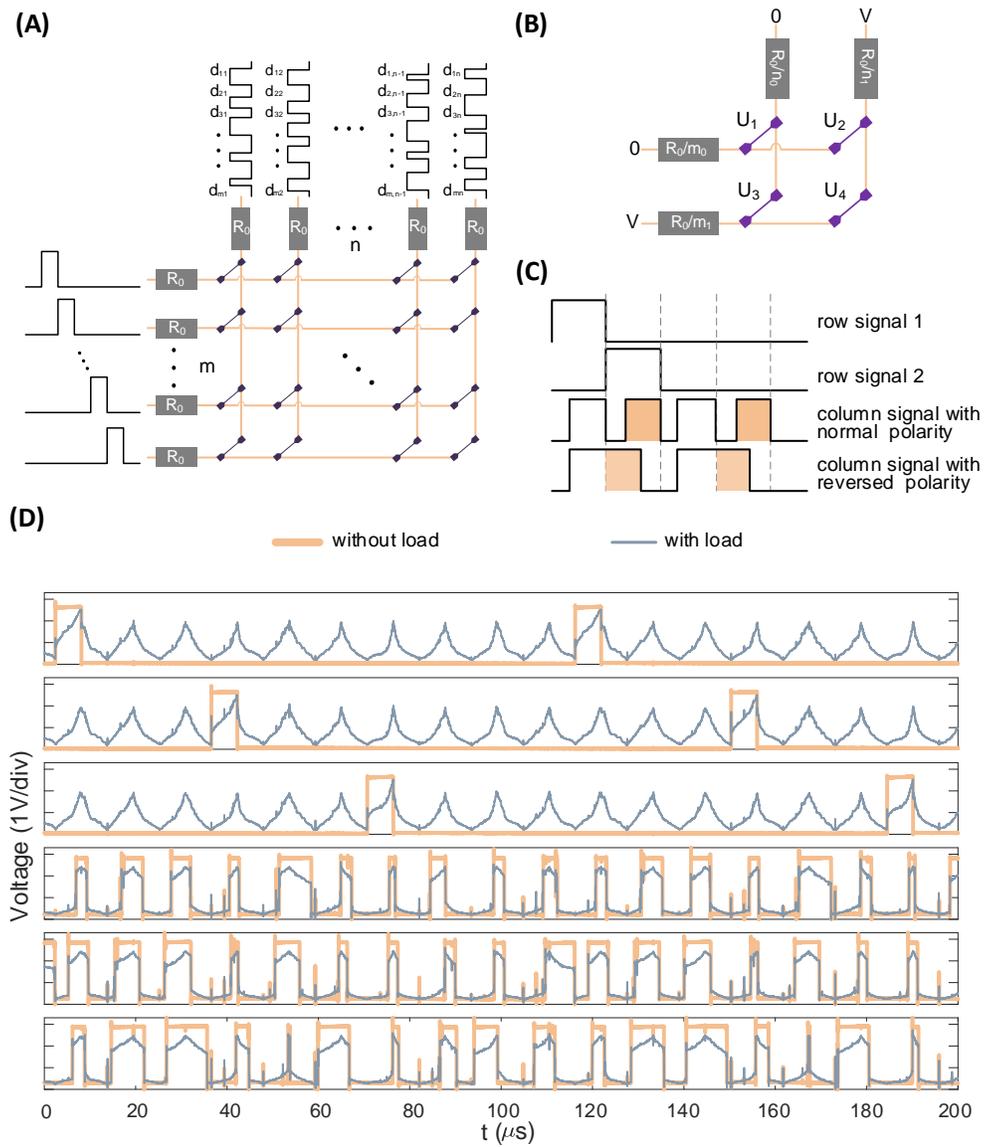

Fig. 3. PWM control. (A) Electrical model of the PWM control circuits. (B) The simplified electrical mode. The input is either high (V) or low (0), the voltages of the four equivalent resistors are ($U_1$, $U_2$, $U_3$, $U_4$). (C) The reversed polarity of the PWM signal in the columns. (D) Measured PWM signals from columns and rows for GPIOs with and without load.

## 4. OPA chip and Beam steering

*4.1 OPA chip fabrication*

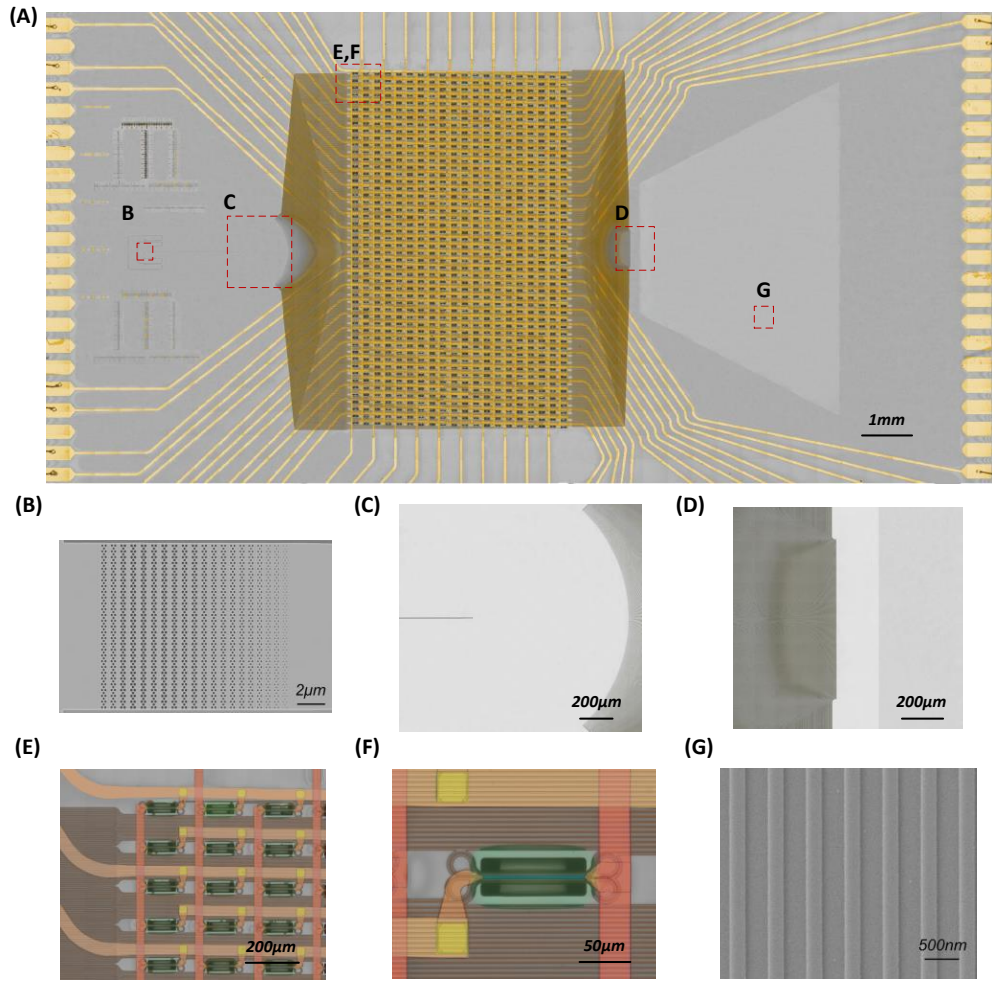

Fig. 4. Chip image. (A) Microscope image of the OPA chip. (B) Grating coupler. (C) SEM of star coupler. (D) Waveguide array. (E) Phase shifter array including column metal lines, row metal lines, heaters, trenches and vias are illustrated by fake color. (F) Single phase shifter. (G) SEM of slab grating.

The OPA device is fabricated on a silicon on insulator wafer with a 220 nm thick silicon layer and a 2 µm bottom oxide layer. The input grating coupler [30], waveguides, star coupler, and phase shifters were defined by E-beam lithography followed by a deep etching process. The slab grating was defined by a second alignment E-beam lithography followed by a 10 nm shallow etch. A 1 µm thick silicon oxide layer was deposited using PECVD to isolate the silicon layer from the metal layer. A 100 nm thick titanium layer and a 5 nm gold layer were deposited and patterned by lift-off to form the heaters on top of the phase shifter region. A 300 nm thick gold layer was deposited and lifted-off to serve as the first electric routing layer, connecting the heaters. Another 1 µm thick silicon oxide layer was then deposited as an isolation layer, on which vias were open and filled by titanium and gold. A 600 nm thick gold layer was deposited and lifted off to serve as the second routing layer, and connected to the first routing layer through the vias. Deep trenches were etched around the heaters through the bottom oxide layer to the silicon substrate, suspending the phase shifter region. The chip was mounted onto a PCB and connected to the external electronics by wire bonding. The microscope image of the fabricated OPA chip is shown in Fig. 4(A), with key components including grating coupler, star coupler, phase shifters, and waveguide array highlighted in the zoomed-in SEM images (Fig. 4B, 4G) and microscope images (Fig. 4C-4F).

*4.2 OPA chip characterization*

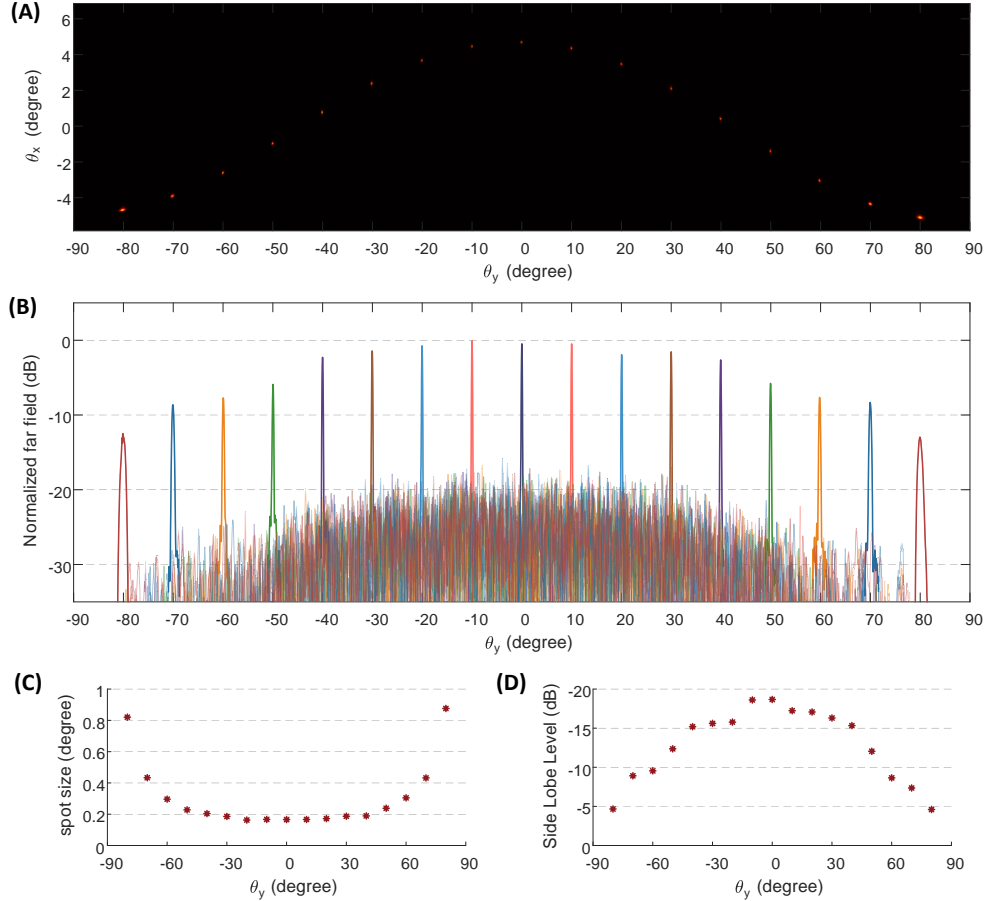

Fig. 5. Far-field emission pattern. (A) Stitched image of the far-field pattern for the beam steered from −80° to 80° at 10° interval in the horizontal direction. (B) Measured average far-field intensity at angle from −80° to 80° along the horizontal direction over the entire 180° FOV. (C) Spot size of each angle from −80° to 80°. (D) SLL of each angle from −80° to 80°.

The OPA device, wire-bonded to a PCB, was mounted on a 3-axis moving stage placed at the center of a rotation stage controlled by a stepper motor. Two cameras (Hamamatsu and Xenics) were mounted at the edge of the rotation stage, both aligned to face the center point of the OPA device. A Fourier lens (Thorlabs) was mounted on the Xenics cameras (Xeva 320) to capture the far-field pattern of the OPA. Each pixel of Xeva 320 camera has a resolvable 4096 intensity points for far field recording. A cylindered convex lens combined with a Fourier lens were mounted on Hamamatsu line-scan camera (C15333-10E) to optimize the far-field beam due to its high frame rate. The far-field image was recorded at 5° intervals by rotating the camera and later stitched together [20] to obtain the full 180° FOV. The PWM control signals were generated by a Xilinx FPGA developer board, which was connected to the PCB through a flat cable. The FPGA board utilized 70 IOs for generating signals for both columns and rows. It was connected to a computer, where an optimization algorithm was performed. Due to variations in optical path lengths and fabrication-induced waveguide width fluctuations, phase errors need to be corrected for each channel. The phase calibration is achieved by optimizing the duty cycles of the PWM signals to maximize the main-lobe power.

Fig. 5(A) shows the composed 2D far-field image of the beam steered from -80° to 80° at 10° interval with the input wavelength at 1550 nm, forming a curved scanning trajectory. This curved trajectory is inherent to slab grating based OPA, which is induced by the coupling of horizontal and vertical wave vector in near field [20]. Further investigation needs to be done to improve the curvature. the  By taking the average of the 2D far field along the vertical axis of Fig. 5(A), we obtained the 1D far-field distribution of the beam at angles ranging from -80° to 80°, as illustrated in Fig. 5(B). As expected, the beam intensity decreases from the center to the edge angles, with an intensity variation of around 13 dB between 0° and ±80°. The noise floor follows a similar envelop, therefore the signal-to-noise ratio is kept almost constant. The beam spot size and side-lobe level (SLL) in the horizontal direction were extracted from Fig. 5(B) for various angles, as shown in Fig. 5(C, D). The beam size at 0° is measured to be 0.17°. It is worth noting that the beam size is slightly increased due to Gaussian amplitude distribution, which reduces the effective size of the aperture. The lowest SLL of -18.7 dB is achieved at 0°, while the SLL is increased to -0.9 dB when the beam is steered to 85° (as detailed in the supplementary material). The SLL remains below -15.2 dB within the ±40° range. Notably, no grating lobes are present across the entire 180° FOV at any beam steering angle. However, when the input wavelength is changed to 1530 nm—where half the wavelength is smaller than the 775 nm pitch—grating lobe appears when the beam is steered to 80° (supplementary).

Beyond the 180° FOV, the absence of grating lobes results in a higher concentration of power in the main lobe compared to traditional waveguide grating based OPA. At a steering angle of 0°, the ratio of power in the main lobe to the total emitted power is measured to be 72%. This demonstrates efficient power concentration in the main lobe, significantly increasing main-lobe emission efficiency and reducing undesirable background noise.

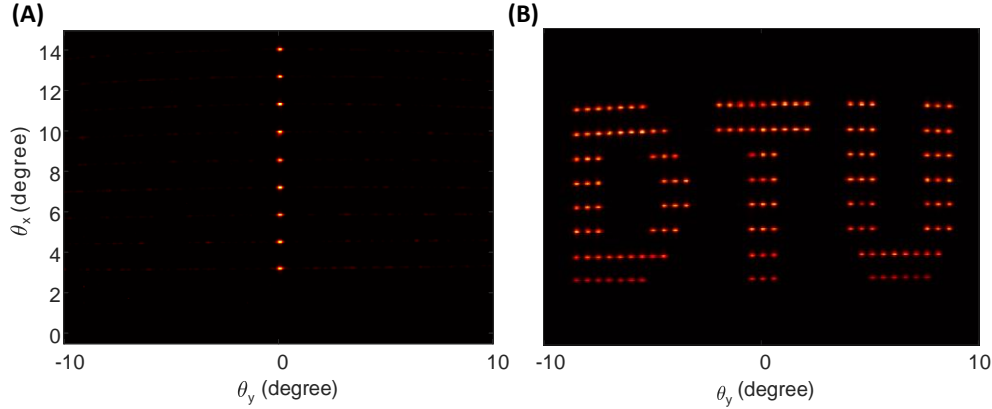

Fig. 6. 2D beam steering.  (A) Beam steering along vertical direction by wavelength tuning. (B) Arbitrary 2D beam steering to form a "DTU" pattern.

For 2D beam steering, the input wavelength is tuned from 1500 nm to 1580 nm, achieving vertical beam steering over a range of 10.8° with a vertical spot size of 0.07°. The measured beam profile is shown in Fig. 6(A). By combining wavelength tuning for vertical beam steering and phase tuning for horizontal beam steering, full 2D beam steering can be realized. To demonstrate this, a "DTU" pattern composed of discrete beam spots is generated using 2D beam steering, as shown in Fig. 6(B). Beam steering is performed in slow speed for camera recording. Fast scanning has not been done due to electronic driver, but we expect the speed to be comparable with the response of phase shifters which is around 3-5 kHz.

*4.3 Chip loss evaluation*

The total chip loss includes the loss from grating coupler, star coupler, waveguide scattering, phase shifters and grating emission. The losses except for star coupler and grating emission are not unique to the slab grating based OPA we proposed in this work. They are estimated to be around 9dB, including 6dB for not optimized grating coupler, 3 dB for waveguide and phase shifters loss in our device. The loss of the star coupler is simulated to be 1.7 dB as presented in supplementary section 5. It mainly comes from the mode mismatch at the fan-out waveguides, and the incomplete capture of diffracted fields from the input waveguides by the fan-out waveguides. The loss of the slab grating emitter also comes from two parts, the downward emission and grating lobes in the near field. The simulated result is shown in the supplementary section 7. At center angle of 0°, the downward emission and grating lobes in the near field induced loss is 2.5 dB and 1 dB, respectively, thus the total insertion loss is estimated to be ~14dB. The downward emission can be improved by grating structure and bottom oxide thickness optimization. The loss to grating lobe in near field can also be improved by optimizing element factor thus less power is dissipated into grating lobes in near field.

## 5. Conclusion

We have demonstrated a highly scalable OPA architecture that simultaneously achieves a wide FOV, high resolution and high beam quality. Compared to our previous 64-channel OPA with each phase shifter controlled by one DAC, we use a row-column PWM control scheme to successfully enable arbitrary phase control of 1000 channels using only 70 electrical signals, substantially increase the channel scale and meanwhile significantly reducing controlling complexity, packaging challenges, and cost. This control scheme can be broadly applicable to a wide range of large-scale photonic integrated circuits beyond OPAs, offering a scalable and cost-effective solution for next-generation programmable photonic systems. Furthermore, by combining a slab grating with a half-wavelength-pitch waveguide array at a large scale, we have achieved grating-lobe-free beam steering over a full 180° FOV, with a high beam quality demonstrated by a minimum SLL of -18.7 dB at 0°. Our results validate the feasibility of large-scale OPAs with a high resolution while maintaining 180° FOV, providing a scalable path forward for high-performance beam steering systems.

**Funding.** Novo Nordisk Foundation (NNF22OC0080333); Novo Nordisk Foundation (NNF23OC0086761); Independent Research Fund Denmark (3105-00367B)

**Disclosures.** YL, HH (P).

**Data availability.** Data may be obtained from the authors upon reasonable request.

**Supplemental document.** See Supplement 1 for supporting content.

# Supplementary 1

## 1. PWM controlling

In the simplified circuits in Fig. 3(B), the four resistors $(R_1, R_2, R_3, R_4)$ with corresponding voltages $(U_1, U_2, U_3, U_4)$ are expressed as:

$$\begin{bmatrix} R_1 \\ R_2 \\ R_3 \\ R_4 \end{bmatrix} = \begin{bmatrix} R/m_0 n_0 \\ R/m_0 n_1 \\ R/m_1 n_0 \\ R/m_1 n_1 \end{bmatrix} \tag{1}$$

where $m_0$ is count of ports that are grounded in rows, $n_0$ is the count of ports that are grounded in columns, $m_1$ is count of ports that are high in rows, and $n_1$ is count of ports that are high in columns, $m_0 + m_1 = m$ and $n_0 + n_1 = n$. The relationship between the current $(I_1, I_2, I_3, I_4)$ through these resistors $(R_1, R_2, R_3, R_4)$ and input voltage $(0, V, V, 0)$ applied at the ports of the equivalent circuit can thus be written as:

$$\begin{bmatrix} 0 \\ V \\ V \\ 0 \end{bmatrix} = A \begin{bmatrix} I_1 \\ I_2 \\ I_3 \\ I_4 \end{bmatrix} \tag{2}$$

where $A$ is the coefficient matrix determined by the port counts $(m_0, m_1, n_0, n_1)$ and resistances as below.

$$\begin{bmatrix} (m_0 + n_0) + R/R_0 & m_0 & n_0 & 0 \\ R_0/n_0 & R_0((m_1 + n_0) + R/R_0)/m_1 & 0 & R_0/m_1 \\ R_0/n_0 & 0 & R_0((m_0 + n_1) + R/R_0)/m_0 n_1 & R_0/n_1 \\ 0 & n_1 & m_1 & (m_1 + n_1) + R/R_0 \end{bmatrix}$$

Solving Eq. (2) gives the node voltages across the equivalent resistors:

$$\begin{bmatrix} U_1 \\ U_2 \\ U_3 \\ U_4 \end{bmatrix} = \begin{bmatrix} I_1 R_1 \\ I_2 R_2 \\ I_3 R_3 \\ I_4 R_4 \end{bmatrix} = B * V \tag{3}$$

where $B$ is a coefficient matrix depending on $(m_0, m_1, n_0, n_1)$ as below.

$$\begin{bmatrix} n_1(R/R_0 + n) - m_1(R/R_0 + m) \\ (R/R_0 + m)(R/R_0 + n) + m_0(R/R_0 + m) + n_1(R/R_0 + n) \\ -((R/R_0 + m)(R/R_0 + n) + m_1(R/R_0 + m) + n_0(R/R_0 + n)) \\ -(n_0(R/R_0 + n) - m_0(R/R_0 + m)) \end{bmatrix} * \frac{R/R_0}{(R/R_0 + m)(R/R_0 + n)(R/R_0 + m + n)}$$

These voltages are fully determined by the number of row and column ports in the grounded (0) and high (V) states.

In our implementation, only one row is high at a time, i.e. $m_0 = m - 1, m_1 = 1$. Therefore, only the column states $(n_0, n_1)$ need to be evaluated. The four voltages only depend on $n_0$ (or $n_1$) and are noted as $U_k^{n_0 = j}$. Since each resistor are cycling within these four voltage states with different $n_0$, we need to find out the ratio of the time to one control period spent on these voltages. Below is the derivation.

For each time slot of the column PWM signals [Fig. S1 (A, B)], the duty cycles from all $n$ columns are sorted in descending order [Fig. S1(C)]. This produces a rank sequence $d_{i[1]} \leq$

$d_{i[2]} \leq \cdots \leq d_{i[n]}$ (bracket index $[j]$ denotes "sorted order"). The differences between successive terms,

$$\Delta d_{i[j]} = d_{i[j]} - d_{i[j-1]}, \quad [j] = 1, \ldots, n, n+1 \quad d_{i[0]} = 0; d_{i[n+1]} = 1$$

specify the fraction of time spent in each configuration. In interval $\Delta d_{i[j]}$, exactly $r_{ij} = [j] - 1$ ($n_0 = r_{ij}$) columns are grounded. For example, in interval $\Delta d_{i[1]}$, 0 columns are grounded. Thus:

- $r_{ij}$ represents the number of grounded columns ($n_0$) in the selected slot,
- $\Delta d_{i[j]}$ represents the duration spent in that configuration.
- Below the rank $r_{ij}$, during the period from 0 to $d_{i[j]}$, the corresponding resistor is in the state of $U_2$ or $U_4$, which mean the column signals are high. Above the rank $r_{ij}$, during the period from $d_{i[j]}$ to 1, the corresponding resistor is in the state of $U_1$ or $U_3$, which mean the column signals are grounded.

The effective power delivered to each resistor is therefore:

$$P_{ij} = \frac{1}{mR} \left( \sum_{[j]=1}^{r_{ij}} \Delta d_{i[j]} \left(U_4^{n_0=[j]-1}\right)^2 + \sum_{[j]=r_{ij}+1}^{n} \Delta d_{i[j]} \left(U_3^{n_0=[j]-1}\right)^2 \right. \\ \left. + \sum_{k=1, k \neq i}^{m} \left( \sum_{[j]=1}^{r_{kj}} \Delta d_{k[j]} \left(U_2^{n_0=[j]-1}\right)^2 + \sum_{[j]=r_{kj}+1}^{n} \Delta d_{k[j]} \left(U_1^{n_0=[j]-1}\right)^2 \right) \right) \quad (4)$$

Finally, Eq. (4) can be reorganized by grouping terms corresponding to the four possible voltage states. This yields the compact form presented as Eq. (1) in the main text,

$$P_{ij} = \frac{1}{mR} \sum_{k=1}^{4} \sum \overrightarrow{\Delta d_k} \cdot \overrightarrow{U_k^2}$$

which is more intuitive. It shows that the effective power at each resistor is governed by all $m \times n$ duty cycles of the column PWM signals. In this way, independent control of the full heater array is achieved.

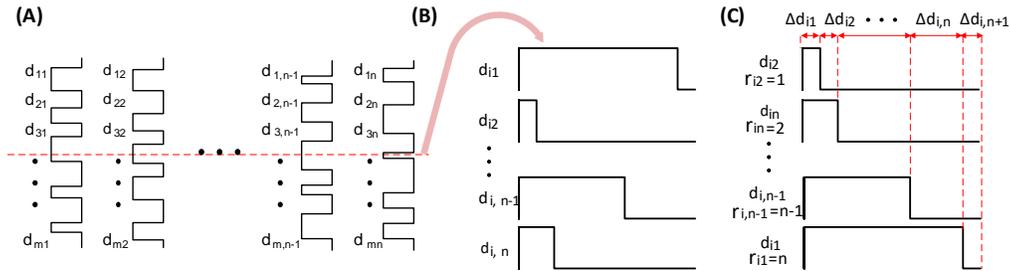

Fig. S1. PWM signals. (A) Example column PWM signals. (B) A single time slot is highlighted. (C) Duty cycles from all $n$ columns are sorted in ascending order. This decomposition allows the column states and corresponding resistor voltages to be evaluated piecewise over time.

2. **Additional Beam steering.**

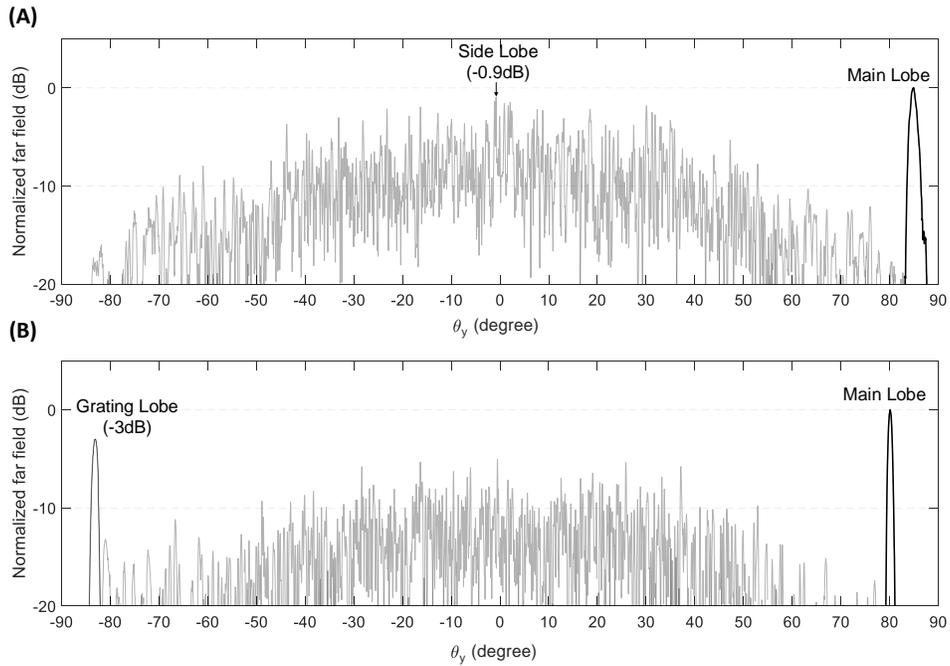

Fig. S2. Additional far field. (A) 1D far field at 85° with input wavelength of 1550nm. SLL is approaching 0dB. (B) 1D far field at 80° with input wavelength of 1530nm. Grating lobe can be observed around angle -83°.

Beam steering beyond ±80° at 1550nm wavelength input is also done in the experiment. The far field of beam steered to 85° is shown in Fig. S2 (A). At this angle, the side lobes in the field center are almost at equal level as the main lobe. Beyond this angle, the side lobe exceeds the main lobe and becomes dominant. Beam steering to 80° at shorter wavelength 1530nm has also been performed. Since 775nm pitch of the waveguide array are larger than half of 1530nm, there are grating lobes observed at around -83° of the field.

## 3. Measurement of phase tuning efficiency.

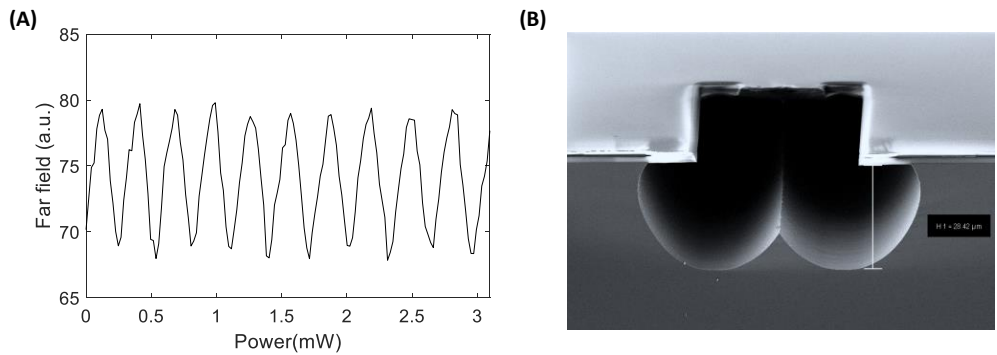

Fig. S3. Phase shifters. (A) Far field intensity versus the applied power to the phase shifters. (B) The undercut of phase shifters, fabricated by isotropic etching of silicon substrate.

Phase tuning efficiency of the phase shifter is measured from another OPA chip with the same phase shifter design. Each phase shifter in this OPA chip is controlled by DAC individually, thus the $\pi$ phase tuning efficiency $P_\pi$ can be measured conveniently. We gradually tune the applied power of a single channel of phase shifter in this OPA by changing the voltage of the

DAC, then record the intensity change of the far field at specific point, as shown in Fig. S3 (A). The $P_\pi$ is measured as half of the cycle in the figure, which is 0.14mW. The low $P_\pi$ attributes to the undercut, which is a process removing the silicon substrate under the phase shifter. It isolates the phase shifters as well as keep the power from dissipating through the substrates. The SEM image of the undercut is shown in Fig. S3 (B).

## 4. Calculation of coupling length

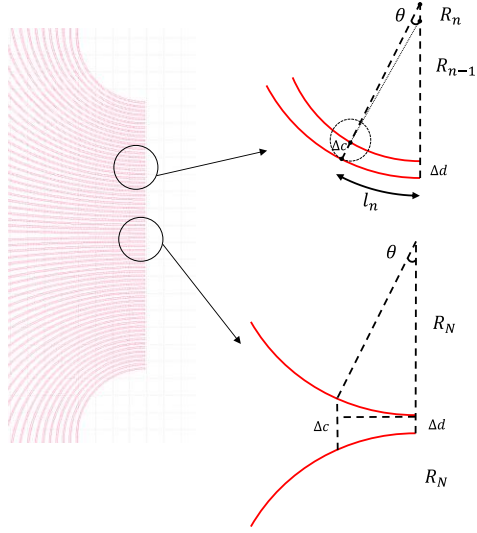

Fig. S4. The geometry of waveguide array.

Fig. S4 shows the geometry of waveguide array. In the center part of the waveguide array, where the waveguides bend in opposite direction (with radius $R_{N/2}$), the length of the center waveguide $l_{N/2}$, where the spacing between the two waveguides increases from the emission pitch $\Delta d$ to distance $\Delta c$, is given by

$$l_{N/2} = R_{N/2} \cos^{-1}\left(\frac{R_{N/2} - (\Delta c - \Delta d)/2}{R_{N/2}}\right) \qquad (5)$$

In the other parts of the waveguide array, where the waveguide bend in the same direction, the equation becomes:

$$l_n = R_n \cos^{-1}\left(\frac{(R_n - \Delta c)^2 + \left(R_n - (R_{n-1} + \Delta d)\right)^2 - R_{n-1}^2}{2(R_n - \Delta c)\left(R_n - (R_{n-1} + \Delta d)\right)}\right) \qquad (6)$$

Where the $R_n, R_{n-1}$ is the bending radius of the two adjacent waveguides, $l_n$ is the length of the waveguides, where the spacing between the two waveguides increases from the emission pitch $\Delta d$ to distance $\Delta c$. Combine these two equations, with the condition of all $l_n$ to be the same, we can calculate the relation between $l_n$ and $N$, which is illustrated in Fig. 2(B).

## 5. Star coupler

The layout of the star coupler is shown in Fig. S5 (A), which consists of an input waveguide, a slab region for beam free propagation and 1000-channel output waveguides. The width of the input waveguide is 400nm, the length of the slab region is 955μm, the width and pitch of output waveguide is 1.3µm and 1.5µm, respectively. The transmissions of center 240 channels are simulated by Lumerical Mode and other channels are fit by Gaussian function, as shown in Fig. S5 (B). The total insertion loss is calculated as -1.7 dB.

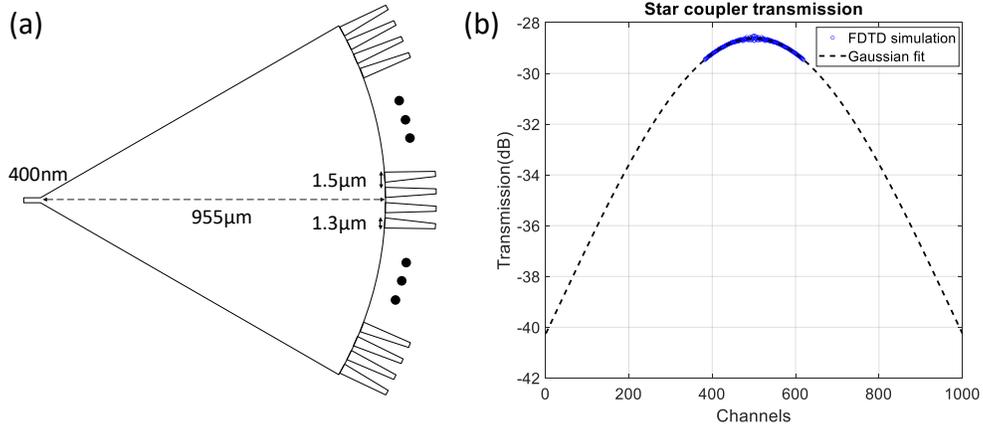

Fig. S5. Phase shifters. (A) The parameters of star coupler. (B) The simulated transmission of star coupler by FDTD and Gaussian fit.

## 6. Simulation for crosstalk in waveguide array

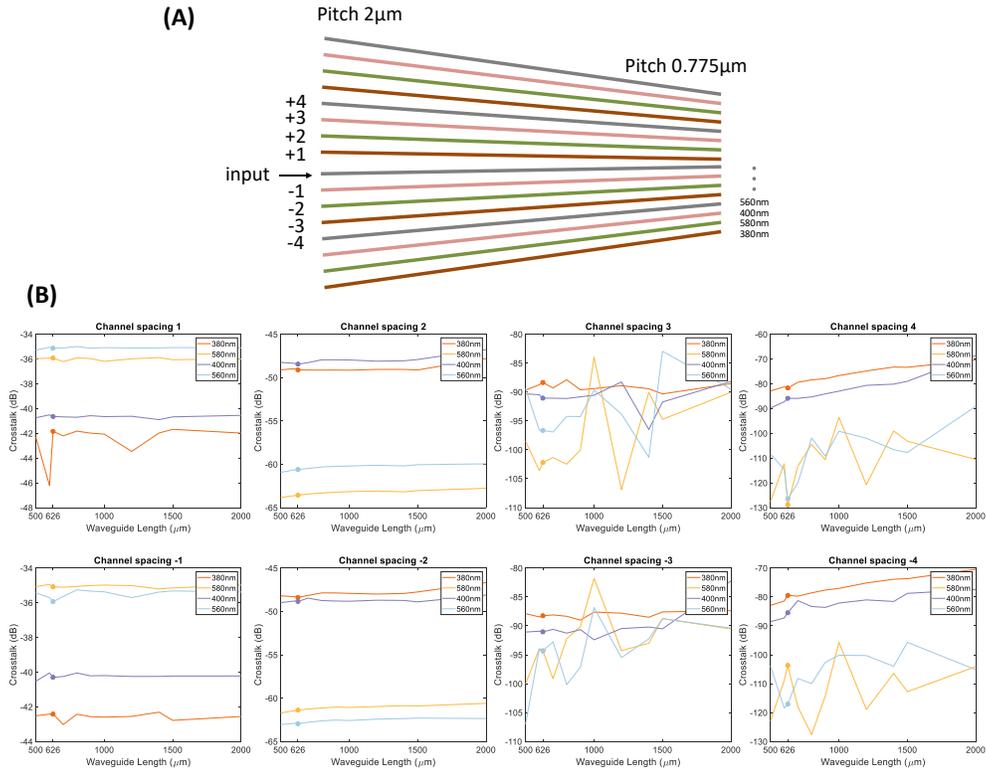

Fig. S6. Simulations of crosstalk in waveguide array. (A) The simulation model of waveguide array, which consist of 4 supperlattice cells. (B) The crosstalk between waveguides with ±1, ±2, ±3, ±4 spacing.

The simulation for waveguide array was done in Lumerical Mode using 2.5D FDTD, with a model of 4 supperlattices, each of which is a four-width waveguide cell. The 16 waveguides array is routing from a spacing of 2um to 0.775nm as shown in Fig. S6 (A). In the simulations, only one waveguide in the center supperlattice cell is connected with a laser source at wavelength 1550nm, and the transmissions in all 16 waveguides are recorded. The crosstalk to the nearest waveguides (±1 channel spacing), the second nearest waveguides (±2 channel spacing), the third nearest waveguides (±3 channel spacing) and the fourth nearest waveguides (±4 channel spacing) are shown in Fig. S6 (B). The crosstalk is most significant between the nearest waveguides.

## 7. Loss of slab grating

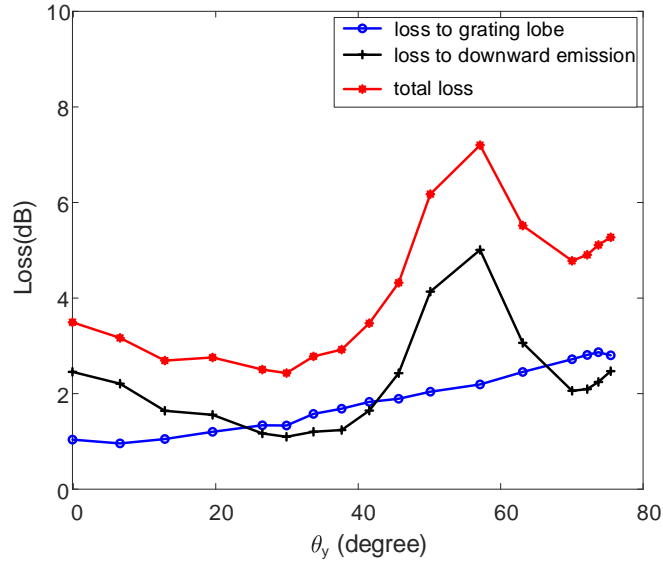

Fig. S7. Simulations of loss of the slab grating, including the loss from grating lobes and downward emission.

The loss of slab grating is caused by the downward emission and the grating lobe in near field. A FDTD simulation was done by a 32-channel waveguides array with 775nm spacing and 450nm width, and a bottom oxide thickness of 2µm. The relative phase difference within the channels were adjusted to achieve beam steering. The power going downward and to grating lobes in near field were recorded as illustrated in Fig. S7. At angle 0°, the total loss is 3.5dB with 1dB from grating lobe loss and 2.5dB from downward emission. At large angles, energy in grating lobes in near field starts to increase, and it either goes downward or propagates outside the slab region. Besides, the ratio between upward emission and downward emission of the main lobe is also angle dependent, because the path length difference caused by propagating through bottom oxide and reflecting by silicon substrate is angle dependent and constructive and destructive interference happens at difference angle.